\documentclass[aps,pra,twocolumn,superscriptaddress,amsmath,amssymb,amsfonts,floatfix]{revtex4}
\usepackage{bbm,graphicx,hyperref,xcolor,times}
\usepackage{soul}
\usepackage{subfigure}
\usepackage{cleveref}
 \usepackage{mathptmx}

\def\Tr{\hbox{Tr}}

\begin{document}

\title{Detection of squeezing by on-chip glass-integrated homodyne analyzer}

\author{Carmen~Porto}
\affiliation{Quantum Technology Lab, Dipartimento di Fisica, 
Universit\`a degli Studi di Milano, I-20133 Milano, Italy}
 
 \author{Davide~Rusca}
 \affiliation{Istituto di Fotonica e Nanotecnologie - 
 Consiglio Nazionale delle Ricerche, p.za Leonardo da Vinci 32, I-20133 Milan, Italy}
 \affiliation{Dipartimento di Fisica, Politecnico di Milano, 
 p.za Leonardo da Vinci 32, I-20133 Milan, Italy}
 
\author{Simone~Cialdi}
\affiliation{Quantum Technology Lab, Dipartimento di Fisica, 
Universit\`a degli Studi di Milano, I-20133 Milano, Italy}
\affiliation{Istituto Nazionale di Fisica Nucleare, Sezione di  Milano, Via Celoria 16, I-20133 Milan, Italy}

\author{Andrea~Crespi}
\affiliation{Istituto di Fotonica e Nanotecnologie - Consiglio Nazionale delle Ricerche, p.za Leonardo da Vinci 32, I-20133 Milan, Italy}
\affiliation{Dipartimento di Fisica, Politecnico di Milano, p.za Leonardo da Vinci 32, I-20133 Milan, Italy}

\author{Roberto~Osellame}
\affiliation{Istituto di Fotonica e Nanotecnologie - Consiglio Nazionale delle Ricerche, p.za Leonardo da Vinci 32, I-20133 Milan, Italy}
\affiliation{Dipartimento di Fisica, Politecnico di Milano, p.za Leonardo da Vinci 32, I-20133 Milan, Italy}

\author{Dario~Tamascelli}
\affiliation{Quantum Technology Lab, Dipartimento di Fisica, Universit\`a degli Studi di Milano, I-20133 Milano, Italy}

\author{Stefano~Olivares}
\affiliation{Quantum Technology Lab, Dipartimento di Fisica, Universit\`a degli Studi di Milano, I-20133 Milano, Italy}
\affiliation{Istituto Nazionale di Fisica Nucleare, Sezione di  Milano, Via Celoria 16, I-20133 Milan, Italy}

\author{Matteo~G.~A.~Paris}
\affiliation{Quantum Technology Lab, Dipartimento di Fisica, Universit\`a degli Studi di Milano, I-20133 Milano, Italy}
\affiliation{Istituto Nazionale di Fisica Nucleare, Sezione di  Milano, Via Celoria 16, I-20133 Milan, Italy}

\begin{abstract}
We design and demonstrate on-chip homodyne detection operating 
in the quantum regime, i.e. able to detect genuine nonclassical features. 
Our setup exploits a  {\em glass-integrated homodyne analyzer} (IHA) 
entirely fabricated by femtosecond laser micromachining. The IHA incorporates on the same chip 
a balanced waveguide beam splitter and a thermo-optic phase shifter, allowing us to 
record homodyne traces at different phases and to perform reliable quantum state 
tomography. In particular, we show that the IHA allows for the detection of nonclassical 
features of continuous-variable quantum states, such as squeezed states.
\end{abstract}
\maketitle

\textit{Introduction --} Balanced homodyne detection \cite{Yuen,Abbas} is the gold standard technique to characterize 
non-classical states of light, e.g. squeezing \cite{Optical homodyne detection, Noise in homodyne detection, HomodyneParis, Measurement of the quantum states of squeezed light,lvov:09}. As such, it is also a pivotal tool in continuous-variable quantum information processing \cite{GaussianQI,oliv:rev}, where information is encoded in the quadratures of the electromagnetic field. Over the last years, several quantum communication as well as quantum metrology protocols have been proposed and demonstrated within this framework \cite{oliv:HD:rec,expCVQKD, BerniAdaptive} and, indeed, continuous-variables approaches are powerful and promising, since they are robust with respect to losses and may achieve deterministic and unconditional operation \cite{Photonic quantum technologies}.

Scaling up the complexity in bulk optics encounters severe limitations, in particular concerning the
optical phase stability and control, which is a crucial requirement for the manipulation of squeezed
states. This makes the adoption of a monolithic integrated platform highly beneficial. Indeed, while
integrated quantum photonics has recently boosted many experimental demonstrations  and
proposals based on discrete-variable systems \cite{DVSilicaonSilicon, sansoniBS, meanyFS,
siliconQP,tama16}, few on-chip experiments have been reported to date with continuous variable systems. Masada et al. demonstrated fundamental operations for the manipulation of squeezed light states within a reconfigurable silica-on-silicon chip \cite{masadaCV}; however, an external piezo-electric controller was used in that case to vary the local-oscillator phase in the homodyne measurement.
Very recently, the integration on the same chip of a beam splitter and balanced detectors of the homodyne apparatus was reported, to perform quantum random number generation \cite{Raffaelli}, but no active modulation was operated on the phase of the local oscillator in the experiment. A few waveguide-based sources of squeezed light states have been also demonstrated \cite{Dutt2015, Stefszky2017} and first steps are moving towards a fully guided-wave based architecture exploiting squeezed light \cite{Kaiser2016}.

In this Letter we report on homodyne measurements performed via an 
integrated-optics device, which is entirely fabricated by femtosecond laser micromachining. 
The device incorporates in the same chip both a balanced waveguide beam splitter and a 
thermo-optic phase shifter. We will refer to this device as {\it integrated 
homodyne analyzer} (IHA). We demonstrate that by means of the thermo-optic phase shifter (TOPS)
it is possible to acquire reliable homodyne traces by scanning different phases of the local 
oscillator and thus  detect nonclassical features of continuous-variable quantum states. More in detail, we investigate the output of the IHA in the presence of coherent and squeezed states and compare the results with those obtained via a standard homodyne detection measurement (SHD), based on a balanced cube beam splitter and a mechanical piezo movement.

\begin{figure}
\includegraphics[width=0.8\columnwidth]{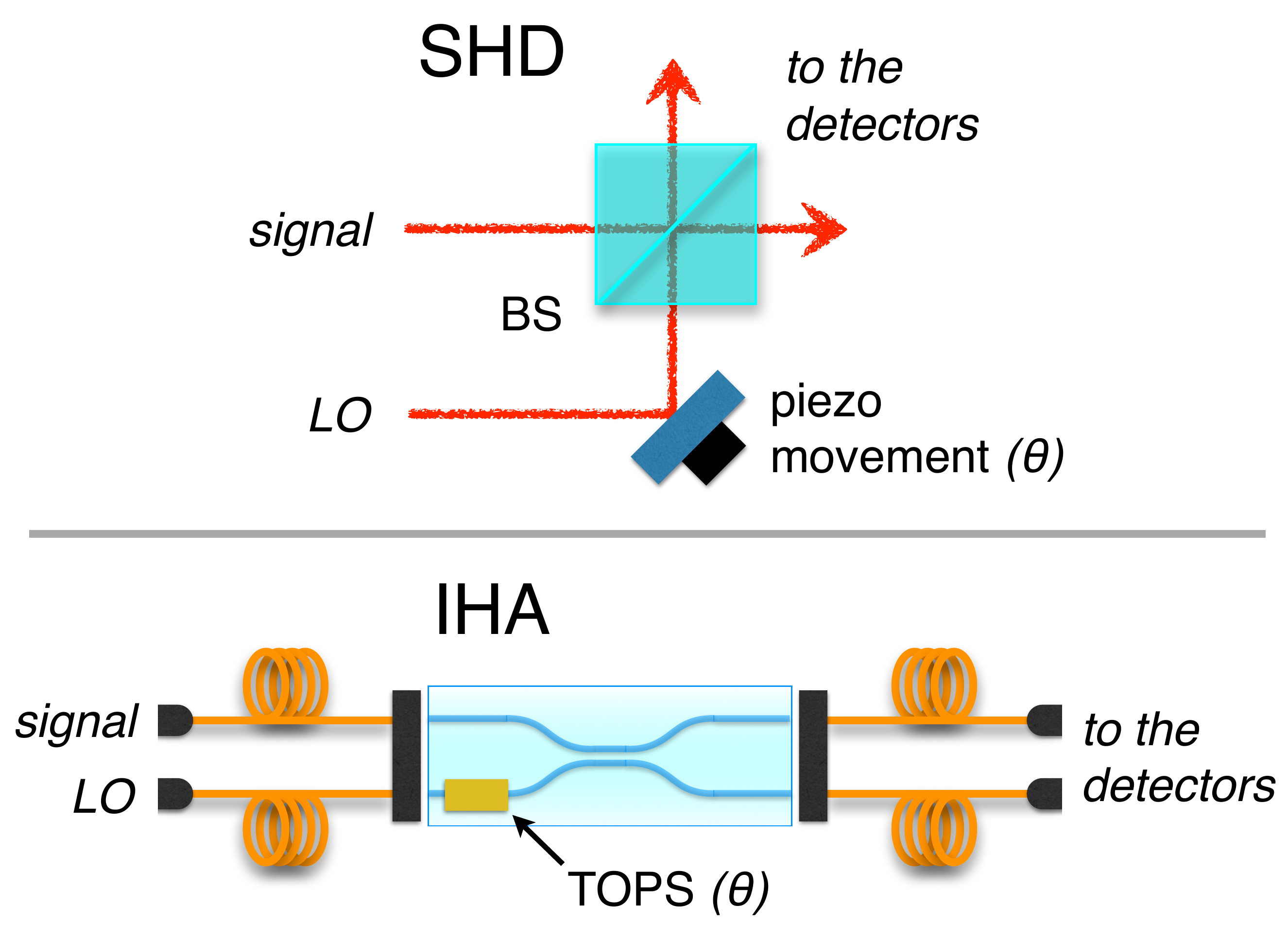}
\vspace{-0.3cm}
\caption[schema]{(Top) The standard homodyne (SHD) setup: signal and LO are mixed at a 50:50 BS,
the quadrature phase $\theta$ is scanned by means of a piezo movement. (Bottom) The integrated homodyne analyzer (IHA):
the BS is replaced by an integrated BS and the phase $\theta$ is scanned by using a thermo-optic phase shifter (TOPS).
Waveguides and fiber arrays are used to couple the input and output channels to the IHA. See the text for details.}
\label{fig:schema}
\end{figure}

\textit{The SHD and the IHA} -- In a typical balanced homodyne measurement \cite{bachor},
the signal field has to be combined with a local oscillator (LO), namely a highly excited 
coherent state, by means of a 50:50 beam splitter (BS), see the top panel of
Fig.~\ref{fig:schema}. The two output beams are then 
detected by two photodiodes and the difference photocurrent is recorded. 
It is well known, in fact, that a measurement of the difference of detected photocurrents 
is proportional to the generic quadrature of the electromagnetic field $\hat{X}
\left( \theta\right) = \hat{a}\, e^{-i\theta} + \hat{a}^{\dag}\, e^{i\theta}$, 
where $\theta$ is the relative phase between LO and the signal described by the 
bosonic annihilation ($\hat{a}$) and creation ($\hat{a}^{\dag}$) operators
with $[\hat{a}, \hat{a}^{\dag}] = \mathbbm{I}$. Therefore, $\theta$ 
defines the phase of the quadrature under investigation and, upon varying it, 
one can perform the quantum-state tomography of the signal field \cite{PFM}. 
In standard homodyne detection (SHD), a cube BS is typically employed 
and $\theta$ is changed by a piezo-actuated mirror placed on the LO optical 
path, as sketched in the top panel of Fig.~\ref{fig:schema}.
In the case of the IHA, this optical 
arrangement is replaced by a waveguide beam splitter, which includes a 
thermo-optic phase shifter (TOPS), see the bottom panel of Fig.~\ref{fig:schema}.

Waveguides are directly inscribed in EAGLE XG (Corning) glass substrate by femtosecond laser writing technology \cite{FLM, eatonTransition}. To fabricate the waveguides, ultrashort pulses of about 300~fs duration, 240~nJ energy and 1~MHz repetition rate, from a Yb:KYW cavity-dumped femtosecond laser oscillator, are focused 30~$\mu$m beneath the surface of the substrate through a 0.6 NA microscope objective. Non-linear absorption processes of the ultrashort laser pulses generate a permanent refractive index increase, localized in the focal region. Translation of the sample under the laser beam at the constant speed of 20~mm/s allows to directly inscribe the waveguide along the desired path; high precision translation is achieved by computer controlled air-bearing stages (Aerotech FiberGLIDE 3D). The waveguides support a single mode at 1064~nm wavelength (mode diameter $1/e^2$ is about 7~$\mu$m). 

The directional coupler is composed of two waveguides, starting at the relative distance of 125~$\mu$m, and brought close at the distance of 11~$\mu$m for a length of 300~$\mu$m. In such region, the waveguides exchange power by evanescent field, and the interaction length is chosen to achieve a balanced splitting ratio. The bent segments have a curvature radius of 90~mm, which produces negligible additional bending losses. 

To fabricate the dynamic phase shifter (the TOPS), a 55~nm gold layer is sputtered on top of the chip and a resistor is patterned, by femtosecond laser pulses, above one of the input waveguides of the directional coupler. This technique is described in more detail in Ref.~\cite{Thermally}. The resistor is 100~$\mu$m wide and 5~mm long, for a value of resistance of about 100~$\Omega$. Thermal dissipation on the resistor, when driven with a suitable current, creates temperature gradients in the glass, and thus thermo-optic modifications of the refractive indices in the waveguides. Therefore, a differential phase, directly proportional to the dissipated electrical power, can be imposed between the two input waveguides. In the homodyne measurements, the resistor is driven by a ramp generator (RG) to scan the LO phase. 

In our experiments we aim at assessing the performance of the IHA, also in comparison with the SHD.
To this purpose, we employed an apparatus (see the Supplemantal Material \cite{SM} for the description of the whole setup) where the two homodyne detection configurations involving the SHD and the IHA, respectively, can be easily and quickly switched by means of flip mirrors. This experimental apparatus, limitedly to the SHD configuration, had been already validated for the generation and detection of different kinds of squeezed states in past experiments \cite{Sidebands,fidelity}. 

The input signals are generated by using a sub-threshold optical parametric oscillator (OPO), which allows to generate squeezed light (see the Supplemantal Material \cite{SM} and Ref.~\cite{Sidebands} for further details). The SHD configuration employs a cube BS, while the LO phase is scanned thanks to a piezo-mounted mirror, linearly driven by the RG. When, instead, the IHA configuration is adopted, the signal from the OPO and the LO is coupled to the inputs of the IHA by a single-mode fiber array; a multi-mode fiber array is then used to couple the outputs of the IHA to the photodiodes (see Fig.~\ref{fig:schema} and the Supplemental Material \cite{SM}).
It is worth noting that a critical issue in our experiments is the minimization of the back reflections across the entire system. To avoid this situation, which induces instabilities in the OPO, the use of ferrule connector/angled physical contact connectors was found to be crucial. Light from the two connectors is finally detected by two low-noise photodetectors, which are shared between the SHD and IHA. The photocurrent difference is finally acquired and spectrally analyzed (further details are given in the Supplemantal Material \cite{SM} and in Ref.~\cite{Sidebands}).

\begin{figure}[tb]
\vspace{0.3cm}
\centering
\includegraphics[width=0.8\columnwidth]{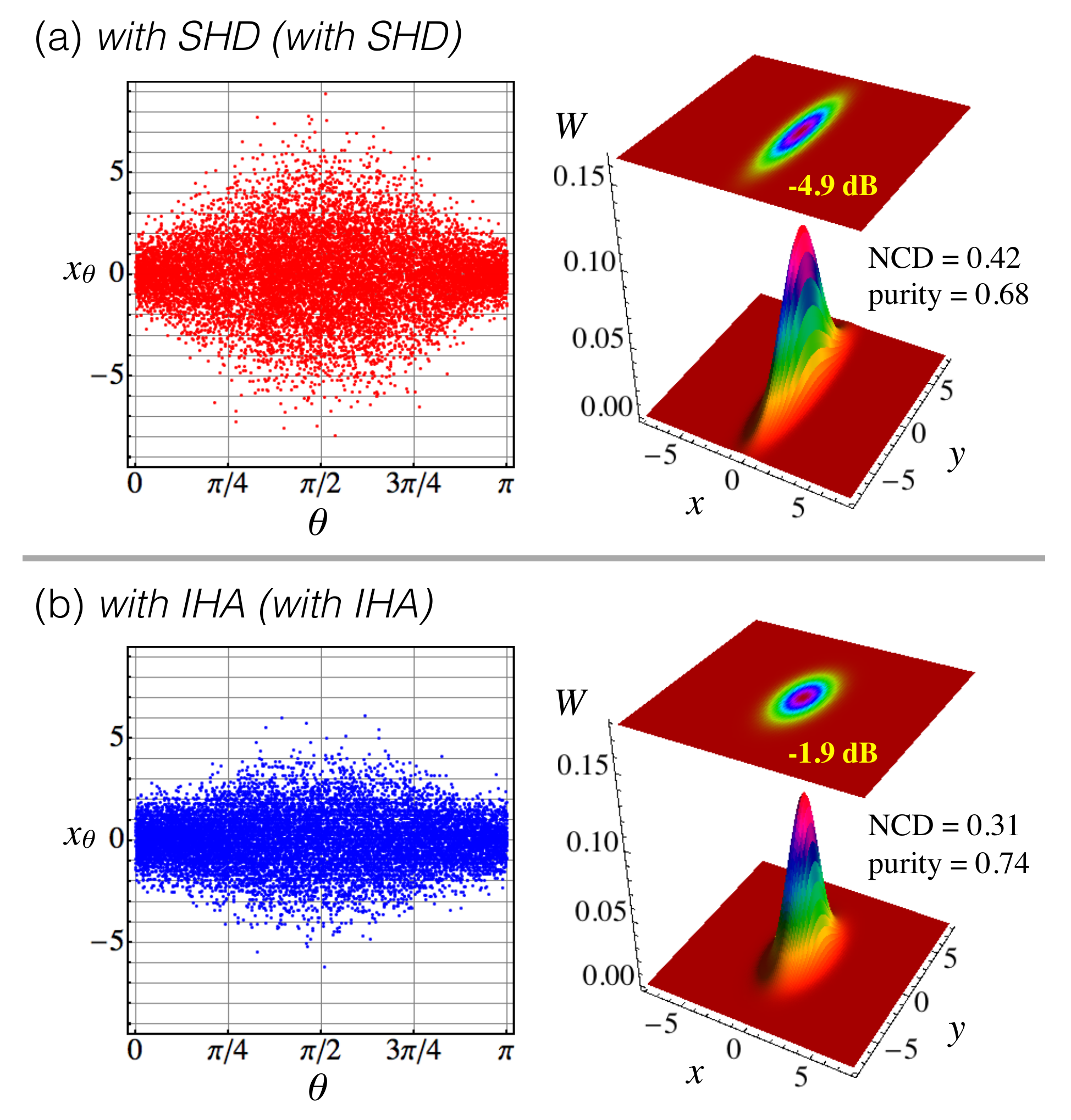}
\vspace{-0.3cm}
\caption[bla]{Homodyne traces referring to the vacuum squeezed states detected by using the BS the SHD (top panel)  and the IHA (bottom panel). We also report the corresponding reconstructed Wigner functions, the nonclassical depth (NCD) and the purity.}
\label{fig:fig2_new}
\end{figure}
\textit{Characterization measurements --}
As a first experiment, we characterize a squeezed vacuum state, generated by the OPO \cite{Sidebands}.
Figure~\ref{fig:fig2_new}  shows the experimental spectral homodyne traces achieved by collecting $M$~=~7000 data points, $\left\lbrace \left( x_k,\theta_k\right) \right\rbrace $, $x_k$ being the $k$-th outcome from the measurement of the quadrature at LO phase $\theta_k$, with $k=1,...,M$. They correspond to the vacuum squeezed states detected by switching between presets. In the top panel of Fig.~\ref{fig:fig2_new} we can see the homodyne trace acquired with the SHD, in which the LO phase is scanned from 0 to $\pi$. The bottom panel of Fig.~\ref{fig:fig2_new}, instead, shows the measurement performed employing the IHA. All traces are normalized to the shot noise level of a pure vacuum field which is therefore set at one. In both cases the pump beam power for the OPO is $P$ = 300~mW (well below the OPO threshold power $P_{\text{thr}}$ = 970~mW , which is obtained by measurement of the classical parametric amplification for the OPO) and the LO power is set to 10~mW by using an amplitude modulator.

As one can see, the traces exhibit squeezing at $\theta=0$ and anti-squeezing at $\theta=\pi/2$.
By applying the pattern function method \cite{PFM}  to $\left\lbrace \left( x_k,\theta_k\right) \right\rbrace $,  we perform a tomographic state reconstruction of single-mode CV systems. The reconstructed Wigner functions which correspond to the two examined cases are displayed next to the respective homodyne traces in Fig.~\ref{fig:fig2_new}. The purity $\mu[\rho] = \Tr[\rho^2]$ and the non classical depth, NCD, \cite{nnClassical} of the state $\rho$ acquired with the SHD are $\mu$ = 0.68 and NCD = 0.42, respectively, whereas those acquired with the IHA are  $\mu$ = 0.74 and NCD = 0.31, respectively. It is worth noting that the higher purity of the state acquired by the IHA is due to the presence of losses, which reduce the state energy and make it nearer to the vacuum state (which is pure). The chip is 20~mm long overall, giving a global insertion loss of about 3~dB at 1064~nm wavelength
(including about 1~dB per facet of estimated coupling loss with the optical fiber).
This is also testified by the squeezing level which is $-4.9$~dB in the case of SHD and is reduced to $-1.9$~dB for the IHA.

\begin{figure}[bt]
\vspace{0.3cm}
\centering
\includegraphics[width=0.75\columnwidth]{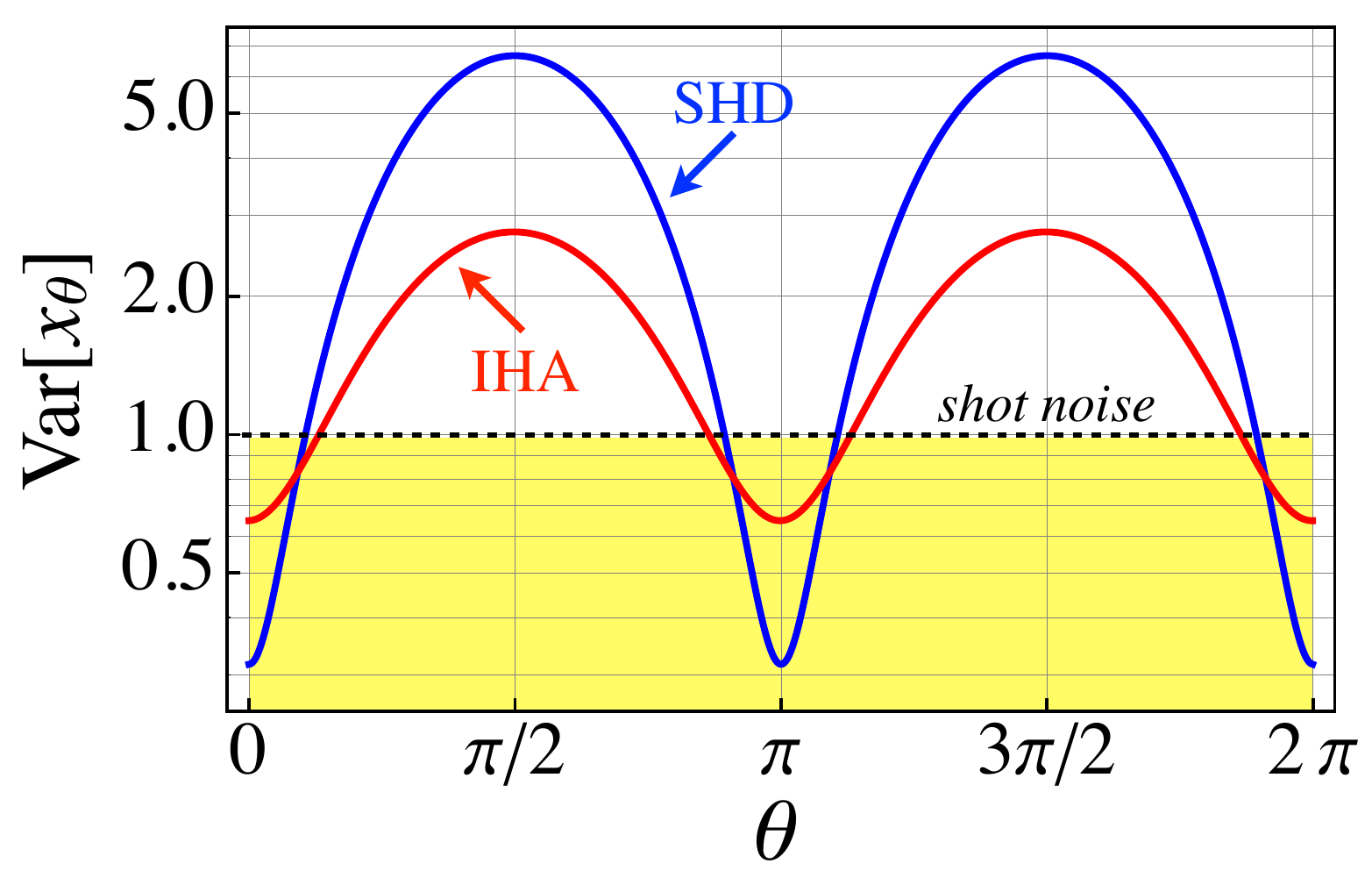}
\vspace{-0.3cm}
\caption{Measured quadrature variance as a function 
of $\theta$ for the SHD (blue) and the IHA (red). We also report 
the shot-noise level for comparison (black horizontal dashed line).}
\label{fig:variances}
\end{figure}

By using pattern function tomography, we can also evaluate the quadrature 
variance Var$[x_\theta]$ as a function of $\theta$ in order to highlight the 
difference between the squeezing and anti-squeezing levels in the two 
measurement configurations.
The results are shown in Fig.~\ref{fig:variances}: the horizontal dashed 
line represents the vacuum noise level and the observed noise levels for 
squeezing are $-4.9 \pm 0.5$~dB for the SHD (blue) and $-1.9 \pm 0.1$~dB for the IHA (red).

This difference arises from the different efficiencies of the two systems. 
Indeed, squeezed states that are observed in practical experiments
necessarily suffer from losses present in transmission channels and detectors 
which degrade the observed squeezing and anti-squeezing levels.
In order to analyse in more detail the effect of losses on the single-mode 
squeezed state, we compare the performance of our apparatus with a theoretical 
model. The noise spectrum $\Delta^2 X_{\pm}$ of the squeezed ($-$) and antisqueezed ($+$) 
quadrature variances for an OPO below threshold can be written as \cite{masadaCV, 15 dB, 
Squeezing at 946nm with periodically poled KTiOPO4}:
\begin{equation}
\Delta^2 \hat{X}_{\pm}=1 \pm \eta_{\text{tot}}\frac{4 \sqrt{P/P_{\text{thr}}}}{\left( 1\mp \sqrt{P/P_{\text{thr}}}\right) ^2+4\left(\frac{2\pi f}{\gamma}\right) ^2 }
\label{spettro}
\end{equation}
where $2\pi f/ \gamma = 0.13$, $f$ being the sideband
frequency of the measurement and $\gamma$ is the OPO cavity decay rate.
We also introduced the total system efficiency $\eta_{tot}$ which can be estimated
by quantifying the individual contributions as 
$
\eta_{\text{tot}}=\eta_{\text{DM}}\, \eta_{\text{esc}}\, 
\eta_{\text{HD}}\, \eta_{\text{D}}\, \eta_{\text{el}}\, ,
$
where $\eta_{\text{DM}}=0.96$ is the propagation efficiency
of the optical path in the space between the OPO
output coupler and the BS, the main amount of which comes from the measured dichroic mirror reflectivity, $\eta_{\text{D}}=0.97$ is the quantum efficiency of photodiodes corresponding to the manufacturer specifications, $\eta_{\text{el}}=0.98$ is the photodetectors' electronic noise and $\eta_{\text{esc}}=0.92$ is the OPO escape efficiency. Finally, $\eta_{\text{HD}}$ is the homodyne detection efficiency and, remarkably, its contribution is not the same in the two different configurations. For the SHD setup one has
$
\eta_{\text{HD}}^{\text{(SHD)}}=\eta_{\text{Vis}}^{\text{(SHD)}} \eta_{\text{BS}}^{\text{(SHD)}}
$
where $\eta_{\text{Vis}}^{\text{(SHD)}} \equiv  {\cal V} ^2$, ${\cal V}$ being the visibility, takes into account the degree of mode matching between OPO output mode and local oscillator LO in BS. The estimated visibility ${\cal V}$ in the case of SHD is $0.96$ and it is achieved by direct measurement of the interference signal between the two beam at BS.
Furthermore, we have $\eta_{\text{BS}}^{\text{(SHD)}}=0.999$, which depends on the not-ideal 50:50 splitting ratio of BS.
  
When, instead, the IHA is used, we also have to take account of the overall IHA efficiency $\eta_{\text{IHA}} = \eta_{\text{f}}\, \eta_{\text{w}}$, where $\eta_{\text{f}}$ and $\eta_{\text{w}}$ are the fiber coupling and the the waveguide transmission efficiencies, respectively. The actual value $\eta_{\text{IHA}}$ has been evaluated by measuring the input intensity into the
fiber coupling lens and the output intensity at the multimode fiber exit. Since the measured fibers coupling efficiency is $\eta_{\text{f}}=0.82$, we estimate $\eta_{\text{w}}=0.51$.

It is worth noting that the chip does not alter the features of the generated states but it acts as a lossy channel. Thus, its effect is merely to degrade the observed squeezing level.
Overall, employing the IHA we have
$
\eta_{\text{HD}}^{\text{(IHA)}}=\eta_{\text{Vis}}^{\text{(IHA)}}\, \eta_{\text{BS}}^{\text{(IHA)}}\, \eta_{\text{C}}
$
with now $\eta_{\text{Vis}}^{\text{(IHA)}} = 0.96$ (the estimated visibility is ${\cal V} = 0.98$), and $\eta_{\text{BS}}^{\text{(IHA)}}=0.998$.

\begin{figure}[tb!]
\vspace{0.3cm}
\centering
\includegraphics[width=0.8\columnwidth]{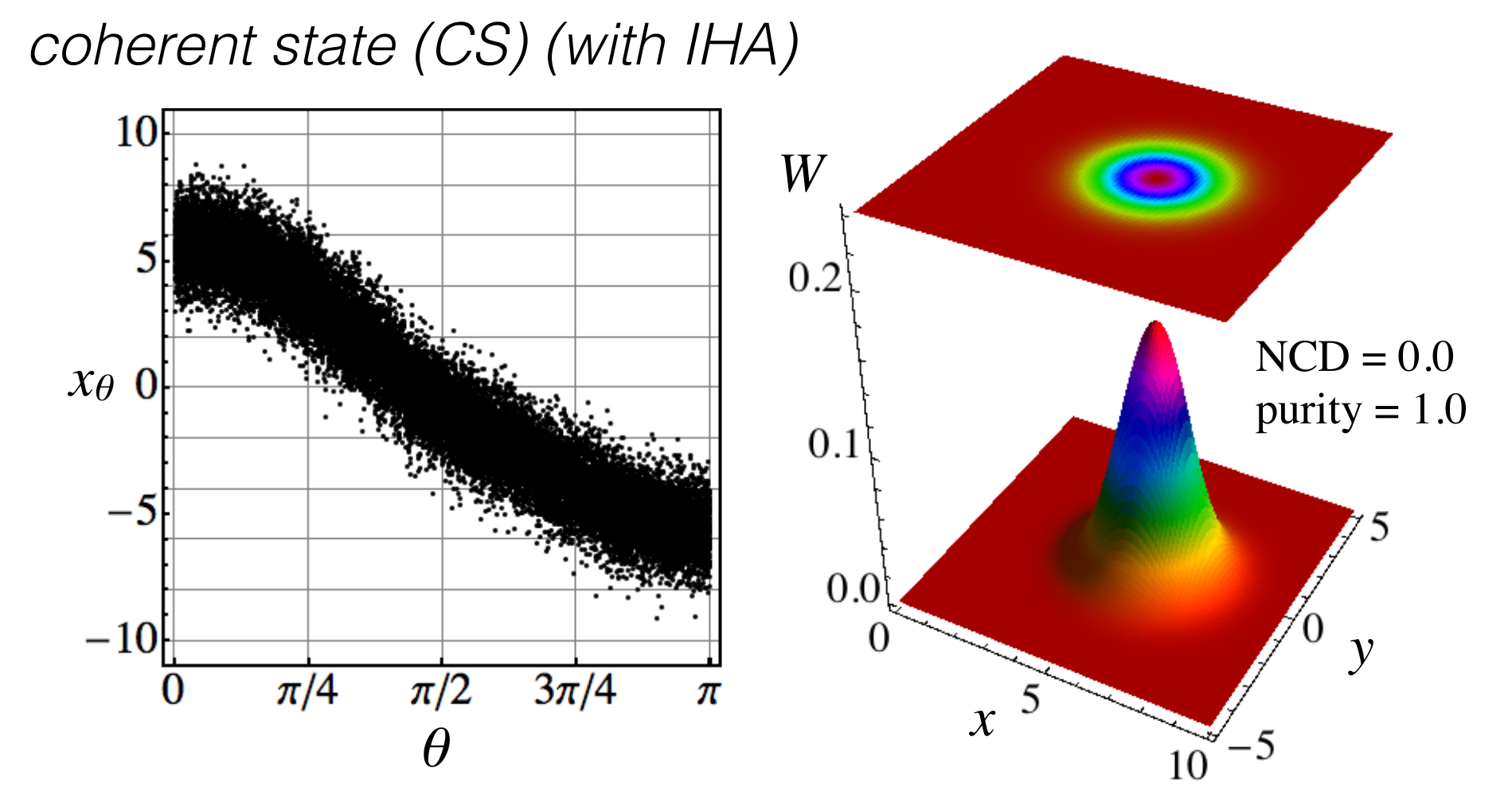}
\vspace{-0.3cm}
\caption{Homodyne trace referring to a coherent state (CS) by exploiting the IHA and the corresponding reconstructed Wigner function. The non classical depth (NCD) and the purity of the state are also reported.}
\label{fig:ch}
\end{figure}
Our experimental setup has the capability to generate different kinds of squeezed states on demand, by injecting states with controlled amplitude and/or phase distributions into the OPO \cite{fidelity}.

In the presence of a coherent state (CS) we obtain the homodyne trace reported in Fig.~\ref{fig:ch}.
We can see that the IHA allows to record the signal at different phases scanned by the thermo-optic phase shifter. Moreover,
the tomographically reconstructed Wigner function correspond to a coherent state with purity 1 and vanishing non classical depth,
as expected.
\begin{figure}[tb!]
\vspace{0.3cm}
\centering
\includegraphics[width=0.8\columnwidth]{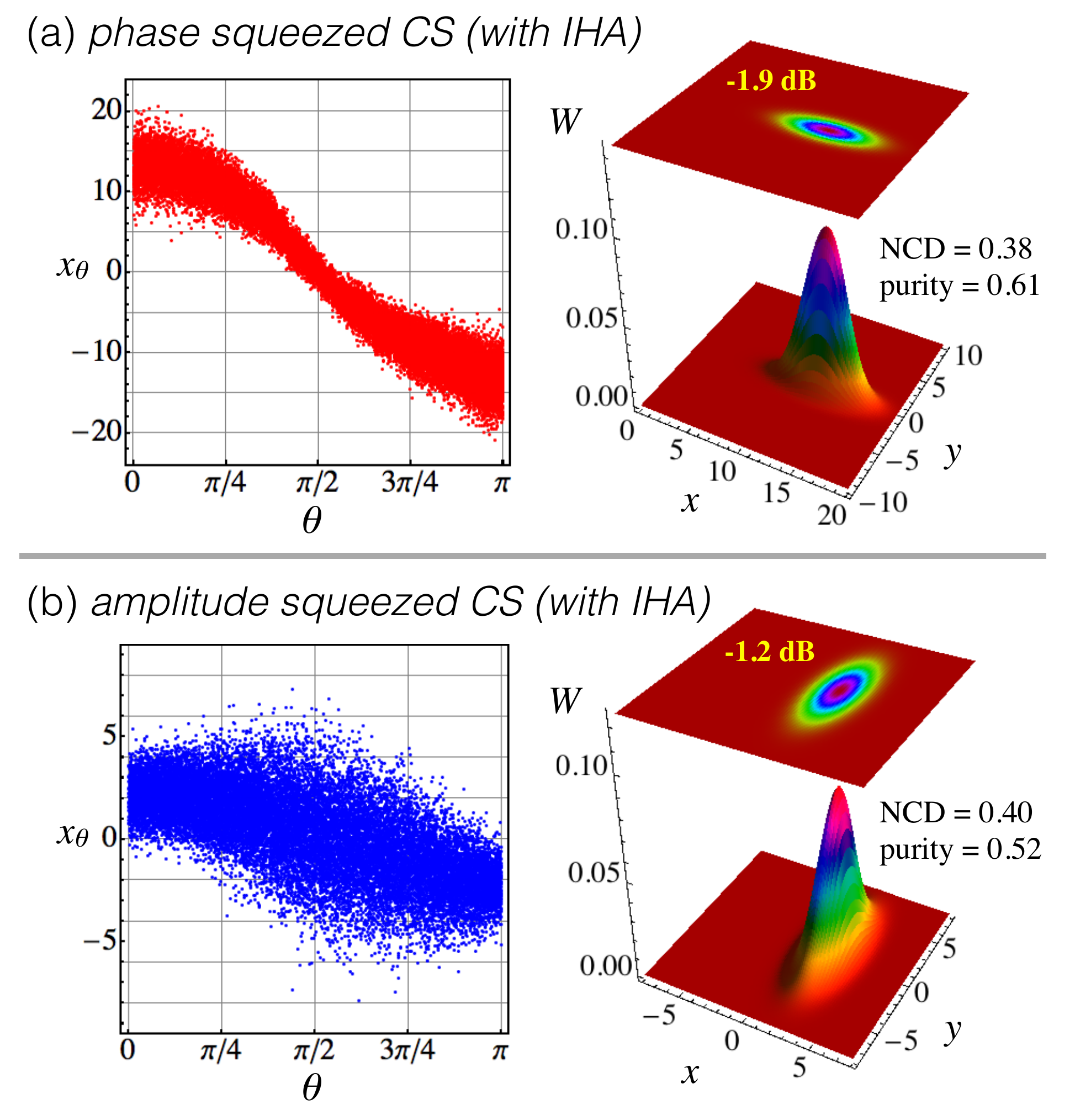}
\vspace{-0.3cm}
\caption{Homodyne traces and reconstructed Wigner functions of phase squeezed CS (top panel) and amplitude squeezed CS (bottom panel) by using the IHA. Their non classical depths (NCD) and purities are also reported.}
\label{fig:ph_amp_sq}
\end{figure}
If the pump and seed fields are in phase, the OPO acts to amplify the seed (phase squeezing); if they are $\pi$ out of phase, it acts to de-amplify the seed (amplitude squeezing). When we operate in amplification regime, we observe the phase squeezed CS shown in top panel of Fig.~\ref{fig:ph_amp_sq}: the observed level for squeezing at $\theta=\pi/2$ is $-1.9 \pm 0.2$~dB. When the OPO is operating to de-amplify the seed (amplitude squeezing), instead, the classical noise on the pump field couples significantly into the OPO output state. This leads to fluctuations in the OPO output signal which significantly degrade the level of squeezing.  The measured amplitude squeezed CS shown in the bottom panel of Fig.~\ref{fig:ph_amp_sq}, indeed features a noise reduction of $-1.2 \pm 0.3$~dB at $\theta=0$. It is worth noting that the difference between the squeezing levels in the two cases shown here would have been the same if we performed the measurements using the SHD.

\textit{Conclusions --} In this letter we have studied in details an integrated-optics device entirely fabricated by femtosecond laser micromachining, the IHA, which incorporates in the same chip both a balanced waveguide beam splitter and a thermo-optic phase shifter. The IHA has been embedded in a typical quantum optics setups for the generation and characterization of continuous-variable optical states, namely, coherent and squeezed-coherent
states. We have demonstrated that, despite the unavoidable losses which affect the IHA,  our integrated device is able to detect the nonclassical features of the input signals.
In particular, the reliability of the recorded homodyne traces have allowed for the tomographic reconstruction
of the considered states by using the pattern function method, thus showing a high degree of reliability also
with respect to the standard homodyne technique based on a cube BS and a piezo-mounted movement to scan the
field quadratures.

Future work will aim at reducing the device losses by further optimization of the waveguide writing process. Anyway, the IHA can already be operated in the quantum regime and embedded in quantum optics setups for the quantum characterization of continuous-variable states.
Our device may be also exploited in hybrid detection schemes, which join
photon number resolving detectors to homodyne-like techniques \cite{bina:17,bina:16}, to implement continuous-variable quantum key distribution protocols based on coherent states
\cite{CV-QKD:01,CV-QKD:03,cattaneo:17}. Our results also open the way to applications of the IHA 
in more elaborated schemes involving continuous-variable optical systems and requiring integrated 
configuration, such as, e.g., the simulation of non Markovian evolution via linear optics 
\cite{strobo:15}.

This work has been supported by the University of Milan through the project ``Continuous-variable quantum technology with integrated quantum photonics'' (fund nr. 15-6-643), and partially supported by EU through the projects QuProCS (grant agreement 641277) and QUCHIP (grant agreement 641039).

\vfill


\clearpage

\section*{Supplemental Material}

\begin{figure*}[tb!]
\includegraphics[width=0.9\textwidth]{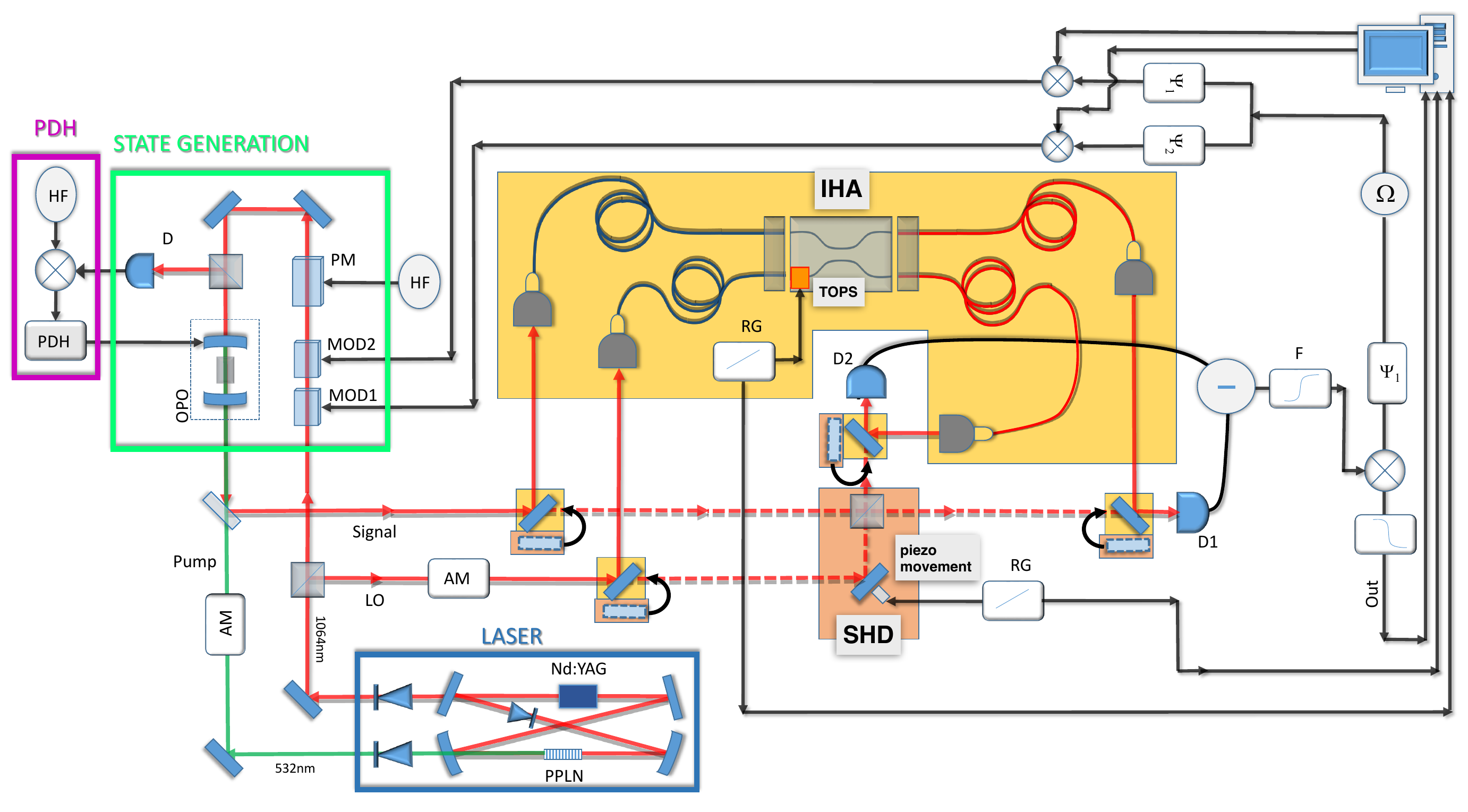}
\vspace{-0.3cm}
\caption[schema]{Experimental setup for standard homodyne detection (SHD) 
and for the integrated homodyne analyzer (IHA). In the latter case, a couple of fiber 
arrays is used for both entrance and exit facets of the IHA to inject and eject signal 
and LO beams efficiently. The LO phase is changed thanks to a thermo-optic phase shifter (TOPS). 
See the text for more detailed informations.}
\label{fig:schema:SM}
\end{figure*}
Figure~\ref{fig:schema:SM} shows the whole scheme of our experimental apparatus.
Continuous-wave squeezed light is generated by a sub-threshold optical parametric oscillator (OPO). The OPO input seed (1064~nm) and the OPO pump beam (532~nm) arise from a home-made internally frequency doubled Nd:YAG laser (Fig.~\ref{fig:schema}). The laser source is based on a four mirrors ring cavity and the active medium is a cylindrical Nd:YAG crystal (diameter 2~mm and length 60~mm) radially pumped by three array of water-cooled laser diodes at 808~nm. The crystal for the frequency doubling is a periodically poled MgO:LiNbO3 (PPLN) of 10~mm, thermally stabilized. A light diode is placed inside the cavity, in order to ensure the single mode operation. The output at 1064~nm is split into two beams by using a polarizing beam splitter (PBS): one is used as the local oscillator (LO) for the homodyne detector and the other is sent into the OPO. The OPO cavity is linear with a free spectral range (FSR) of 3300~MHz, the output mirror has a reflectivity of 91.7\%, and the rear mirror 99.6\%. A phase modulator (PM) generates a signal at frequency of 110~MHz (HF) used as active stabilization of the OPO cavity via the Pound-Drever-Hall (PDH) technique \cite{PDH}. In order to generate the coherent squeezed states our strategy is to exploit the combined effect of two optical modulators placed before the OPO ~\cite{fidelity}. The first modulator generates a coherent state with phase $0$, while the second modulator generates a coherent states with phase $\frac{\pi}{2}$. By matching these coherent states with properly chosen amplitudes, it is possible to generate arbitrary coherent states for seeding the OPO. Once the state is generated, the OPO output beam can be directed into either SHD or IHA configurations, respectively, by means of mirrors with high reflectivity at 1064~nm, mounted on flippers
(see Fig.~\ref{fig:schema}). 

The SHD configuration employs a cube BS, while the LO phase is scanned thanks to a piezo-mounted mirror, linearly driven by the ramp generaor (RG).
When the mirrors are flipped up, the set-up is switched to the IHA configuration. In order to efficiently couple the signal from the OPO and the LO to the inputs of the IHA we use a fiber arrays; a similar fiber arrays is used to couple the outputs of the IHA to the photodiodes, as pictorially shown in Fig.~\ref{fig:schema}. Each fiber array contains two fibers, fastened with high precision at a distance of 125~$\mu$m by means of quartz V-groove blocks. The input fibers, which yield single-mode (SM) and polarization-maintaining operation, are Ferrule Connector/Physical Contact (FC/PC) connectorized. The signal and LO beams are coupled into them by means of adjustable FiberPort micropositioners. To improve the coupling efficiency, the built-in lenses of the latter components are chosen to ensure the optimal match between the incident modes and the fiber modes. 
The integrated-optics chip of the IHA is placed on an aluminium holder, to enhance heat dissipation, and the two fiber arrays can be aligned accurately to the waveguide inputs and outputs by means of two six-axis positioning stages with micrometric resolution. Index-matching gel is used to eliminate reflection losses at the fiber-waveguide interfaces.
The fibers in the output fiber array are instead multimode (MM), with a core diameter of 50~$\mu$m and a cladding diameter of 125~$\mu$m, connectorized at the other ends with ferrule connector/angled physical contact (FC/APC) terminals, to reduce back reflections.

Light from the two FC/APC connectors is finally collected by lenses and focused on low-noise photodetectors (D1, D2), which are shared between the two configurations SHD and IHA. The information about the signal, at a frequency $\Omega$ = 3~MHz, is retrieved by using an electronic apparatus which consists of a mixer ($\otimes$), a phase shifter, set to ensure zero phase shift between the two inputs of the mixer, and a low-pass filter at 300~kHz. The two detectors are based on a Fermionics photodiode and the LMH6624 operation amplifier. The electronic noise is 17~dB below the vacuum noise at 3~MHz for 10~mW of the LO power.

The LO phase is spanned between 0 and 2$\pi$ by using a RG. To perform the homodyne measurements, we collect $M$~=~7000 data points in a time window of 800~ms with a repetition rate of 10~kHz. The sampling is triggered by a signal generated by the RG to ensure the synchronization between the acquisition process and the scanning of LO.

\end{document}